\documentclass[twocolumn,aps,pra,superscriptaddress,showpacs,tightenlines]{revtex4-1}
\usepackage{mathptmx}
\DeclareMathAlphabet{\mathcal}{OMS}{cmsy}{m}{n}
\usepackage{mathrsfs}

\usepackage[T1]{fontenc}
\usepackage[latin9]{inputenc}
\usepackage{amstext}
\usepackage{amsmath}
\usepackage{graphicx}
\usepackage{esint}
\usepackage{color}
\usepackage[colorlinks, linkcolor=blue, citecolor=blue,hyperindex,bookmarks=false,pdfstartview=FitH]{hyperref}
\makeatletter

\@ifundefined{textcolor}{}
{\definecolor{BLACK}{gray}{0}
 \definecolor{WHITE}{gray}{1}
 \definecolor{RED}{rgb}{1,0,0}
 \definecolor{GREEN}{rgb}{0,1,0}
 \definecolor{BLUE}{rgb}{0,0,1}
 \definecolor{CYAN}{cmyk}{1,0,0,0}
 \definecolor{MAGENTA}{cmyk}{0,1,0,0}
 \definecolor{YELLOW}{cmyk}{0,0,1,0}
}

\makeatother
\begin{document}

\title{Energy-level-attraction and heating-resistant-cooling of mechanical resonators with exceptional points}
\author{Cheng Jiang}
\affiliation{School of Physics and Electronic Electrical Engineering, Huaiyin Normal University, Huai'an 223300, China}
\affiliation{Department of Applied Physics, Aalto University, P.O. Box 15100, FI-00076 AALTO, Finland}

\author{Yu-Long Liu}
\email{liuyulonghs@126.com}
\affiliation{Quantum states of matter, Beijing Academy of Quantum Information Sciences, Beijing 100193, China}
\affiliation{Department of Applied Physics, Aalto University, P.O. Box 15100, FI-00076 AALTO, Finland}

\author{Mika A. Sillanp\"{a}\"{a}}
\affiliation{Department of Applied Physics, Aalto University, P.O. Box 15100, FI-00076 AALTO, Finland}

\begin{abstract}
We study the energy-level evolution and ground-state cooling of mechanical resonators under a synthetic phononic gauge field. The tunable gauge phase is mediated by the phase difference between the $\mathcal{PT}$- and anti-$\mathcal{PT}$-symmetric mechanical couplings in a multimode optomechanical system. The transmission spectrum then exhibits the asymmetric Fano line shape or double optomechanically induced transparency by modulating the gauge phase. Moreover, the eigenvalues will collapse and become degenerate although the mechanical coupling is continuously increased. Such counterintuitive energy-attraction, instead of anti-crossing, attributes to destructive interferences between $\mathcal{PT}$- and anti-$\mathcal{PT}$-symmetric couplings. We find that the energy-attraction, as well as the accompanied exceptional points (EPs), can be more intuitively observed in the cavity output power spectrum where the mechanical eigenvalues correspond to the peaks. For mechanical cooling, the average phonon occupation number becomes minimum at these EPs. Especially, phonon transport becomes nonreciprocal and even ideally unidirectional at the EPs. Finally, we propose a heating-resistant ground-state cooling based on the nonreciprocal phonon transport, which is mediated by the gauge field. Towards the quantum regime of macroscopic mechanical resonators, most optomechanical systems are ultimately limited by their intrinsic cavity or mechanical heating. Our work revealed that the thermal energy transfer can be blocked by tuning the gauge phase, which supports a promising route to overpass the notorious heating limitations.
\end{abstract}
\maketitle

\section{Introduction}
Non-Hermitian systems with parity-time ($\mathcal{PT}$) symmetry have attracted considerable attention since the pioneering work of Bender and Boettcher in 1998~\cite{Bender}. The $\mathcal{PT}$-symmetric Hamiltonian $H$, which satisfies the commutation relation  $[H,\hat{P}\hat{T}]=0$ with the $\mathcal{PT}$ operator $\hat{P}\hat{T}$, can exhibit entirely real spectra below some critical system parameters. Moreover, an abrupt phase transition between the unbroken- and broken-symmetry occurs at the exceptional point (EP), where the eigenvalues and the corresponding eigenvectors coalesce. The experimental demonstrations of the $\mathcal{PT}$-symmetry and EP have revealed many intriguing phenomena such as unidirectional transmission~\cite{PengB,ChangL}, single-mode lasers~\cite{FengLScience,HodaeiScience}, and enhanced sensitivity~\cite{Hodaei,ChenWJ}. More comprehensive development related to $\mathcal{PT}$-symmetry can be found in Refs.~\cite{Zyablovsky2014,Konotop2016,Ganainy2017,Makris2018,Miri2019,Yang2019}.

On the other hand, the anti-$\mathcal{PT}$ symmetric Hamiltonian of growing interest satisfies $\{H,\hat{P}\hat{T}\}=0$ and can possess the purely imaginary eigenvalues. Recently, anti-$\mathcal{PT}$ symmetry has been widely observed in atomic systems~\cite{PengP,JiangY}, electrical circuits~\cite{ChoiY}, diffusive thermal materials~\cite{LiYScience}, a magnon-cavity-magnon coupled system~\cite{ZhaoJ}, coupled waveguide systems~\cite{ZhangXL,FanH}, and a single microcavity with nonlinear Brillouin scattering~\cite{ZhangFX}. Such systems can also display some noteworthy effects including constant refraction~\cite{YangF}, nonreciprocity and enhanced sensing~\cite{ZhangHL}, and information flow~\cite{WenJ}. Compared with $\mathcal{PT}$-symmetric systems, anti-$\mathcal{PT}$-symmetric systems don't require any gain medium that may introduce extra instability and experimental complexity~\cite{wujinhui}.

Optomechanical systems, which consist of an electromagnetic cavity coupled with a mechanical resonator via radiation pressure~\cite{Aspelmeyer1,XiongH,LiuYL}, have witnessed significant developments, such as ground state cooling of the mechanical resonator~\cite{Chan,Teufel1,PetersonPRL,ClarkNature,QiuLRPL,DelicScience}, optomechanically induced transparency (OMIT)~\cite{Weis,Naeini1}, and non-classical states of motion~\cite{Wollman,Pirkkalainen}. More recently, multimode optomechanical systems comprising two or more mechanical resonators have been under intensive investigation~\cite{Nielsen,Shkarin,ZhangM,LiWL,Riedinger,KorppiNature,MasselNC,JiangCPB,KorppiPRA,
Weaver,XuH2016,XuH2019,Mathew,YangC}. The displacement of one mechanical resonator changes the cavity resonance and hence the intra-cavity photon number, which will in turn modify the radiation pressure on the other mechanical resonator. Therefore, the mechanical resonators can be coupled indirectly via the common interaction with the cavity field. These systems provide a platform to study synchronization~\cite{ZhangM,LiWL} and entanglement~\cite{Riedinger,KorppiNature} of the mechanical resonators, topological energy transfer~\cite{XuH2016}, nonreciprocal phonon transport~\cite{XuH2019,Mathew}, and so on.

Furthermore, if the mechanical resonators are coupled directly through Coulomb interaction~\cite{Brown,MaPC,ZhangXY}, a piezoelectric
transducer~\cite{Okamoto} or a superconducting charge qubit~\cite{LaiDG1,LaiDG2}, the multimode optomechanical system with loop interaction can exhibit exciting features such as nonreciprocal ground state cooling~\cite{LaiDG1} and enhanced second order sideband~\cite{LaiDG2}. Mechanical $\mathcal{PT}$ symmetry has also been demonstrated in two coupled optomechanical systems with the cavities being driven by blue- and red-detuned laser fields, respectively~\cite{XuXW}. Notable that the direct couplings between mechanical modes are coherent and $\mathcal{PT}$-symmetric. The $\mathcal{PT}$-symmetry broken induced energy localization and ground-state cooling at room temperature have been proposed and detailed discussed in Refs.~\cite{yulong2017,jinghui2017}.

As a counterpart, how anti-$\mathcal{PT}$-symmetry affect the mechanical cooling is essentially intriguing but less discussed. In this paper, we investigate the cooling of the mechanical resonators in a multimode optomechanical system, which consists of two directly coupled mechanical resonators interacting with a common cavity field. When the cavity is driven on the red sideband of the average frequency of the two mechanical resonators, we derive the effective Hamiltonian for the mechanical modes by adiabatically eliminating the cavity field and find that mechanical anti-$\mathcal{PT}$ symmetry can be realized. Especially, when taking both the $\mathcal{PT}$- and anti-$\mathcal{PT}$-symmetric mechanical couplings into the consideration, a phononic gauge field with a tunable phase is synthesized. The EPs at which both the real and imaginary parts of the eigenvalues coalesce periodically appear at the phase-match points. The positions of the EPs can be shifted by modifying the relative strength between $\mathcal{PT}$- and anti-$\mathcal{PT}$-symmetry couplings, which in turn affects the phonon flow and the final phonon occupation numbers. We emphasize that exploring how the phononic gauge field affects the energy-level evolution, as well as the mechanical cooling are the main task of this article.


The paper is organized as follows. In Sec.~\ref{sec:system}, we describe the multimode optomechanical system and then reveal anti-$\mathcal{PT}$-symmetric mechanical couplings mediated by a common cavity field. A phononic gauge field is subsequently constructed and the periodic EPs are also presented in this section. In Sec.~\ref{sec:transmission}, we demonstrate how to observe counterintuitive energy-attraction around the EPs through the transmission and output spectra of the cavity. Then, the gauge phase controlled nonreciprocal phonon transport and heating-resistant cooling of the mechanical resonators are presented in Sec.~\ref{sec:cooling}. Finally, the conclusion of this paper is given in Sec.~\ref{sec:conclusion}.

\section{Phononic gauge field and exceptional points}\label{sec:system}
\begin{figure}
\centering
\includegraphics[width=6cm]{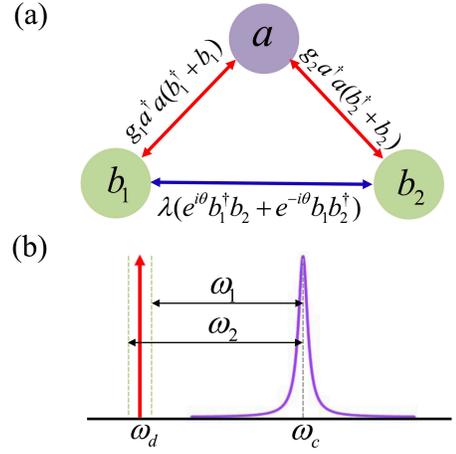}
\caption{(a) Schematic diagram of the optomechanical mechanical system consisting of one cavity mode $a$ and two mechanical modes $b_1$ and $b_2$. The optical mode and the mechanical modes are coupled via radiation pressure, and the two mechanical modes are coupled directly with a phase-dependent coupling strength. (b) The driving scheme in the frequency domain. $\omega_c$ is the frequency of the cavity mode, $\omega_1$ and $\omega_2$ are the frequencies of the two mechanical modes, and $\omega_d$ is the frequency of the driving field, which is red-detuned with respect to the cavity field by the average frequency of the two mechanical modes.}
\label{Fig1}
\end{figure}
We consider the optomechanical system, which consists of two mechanical resonators coupled to a common cavity field via radiation pressure. In addition, the two mechanical resonators are coupled with each other via phase-dependent phonon-exchange interaction~\cite{LaiDG1,LaiDG2}. The Hamiltonian of the system is given by
\begin{eqnarray}
H=&&\hbar\omega_c a^{\dagger}a+\sum_{k=1,2}\hbar\omega_kb_k^{\dagger}b_k+\sum_{k=1,2}\hbar g_ka^{\dagger}a(b_k^{\dagger}+b_k)\nonumber\\&&+\hbar\lambda(e^{i\theta}b_1^\dagger b_2+e^{-i\theta}b_1b_2^\dagger)+i\hbar\sqrt{\kappa_e}\varepsilon_d(a^{\dagger}e^{-i\omega_d t}-a e^{i\omega_d t}), \label{Eq:Hamil}\nonumber\\
\end{eqnarray}
where $a^{\dagger}~(a)$ are the creation (annihilation) operators of the cavity field with resonance frequency $\omega_c$, and $b_k^{\dagger}~(b_k)$ are the creation (annihilation) operators of the mechanical resonators with resonance frequencies $\omega_k~(k=1,2)$. $g_k$ is the single-photon optomechanical strength between the $k$th mechanical resonator and the cavity field. $\lambda$ represents the coupling strength between the two mechanical resonators with the phase $\theta$. The direct interaction between two resonators can be realized by coupling to a superconducting charge qubit~\cite{LaiDG1,LaiDG2}. The last term in Eq. (\ref{Eq:Hamil}) describes the coupling between the driving field at frequency $\omega_d$ and the cavity field, where $\kappa_e$ is the decay rate of the cavity due to external coupling and $\varepsilon_d$ is the amplitude of the driving field. In the rotating frame at the driving frequency $\omega_d$, the Hamiltonian becomes
\begin{eqnarray}
H=&&\protect\hbar\Delta_c a^{\dagger}a+\sum_{k=1,2}\hbar\omega_kb_k^{\dagger}b_k+\sum_{k=1,2}\hbar g_ka^{\dagger}a(b_k^{\dagger}+b_k)\nonumber\\&&+\hbar\lambda(e^{i\theta}b_1^\dagger b_2+e^{-i\theta}b_1 b_2^\dagger)+i\hbar\sqrt{\kappa_e}\varepsilon_d(a^{\dagger}-a),
\end{eqnarray}
where $\Delta_c=\omega_c-\omega_d$ is the detuning between the cavity and the driving field.

The dynamics of the system is determined by the following quantum Langevin equations (QLEs):
\begin{eqnarray}
\dot{a}=&&-\left(\frac{\kappa}{2}+i\Delta_c\right)a-ig_1(b_1^{\dagger}+b_1)a-ig_2(b_2^{\dagger}+b_2)a+
\sqrt{\kappa_e}\varepsilon_L\nonumber\\&&+\sqrt{\kappa_e}a_{\mathrm{in},e}+\sqrt{\kappa_i}a_{\mathrm{in},i}, \label{Eq:da}\\
\dot{b_1}=&&-\left(\frac{\gamma_1}{2}+i\omega_1\right)b_1-i\lambda e^{i\theta}b_2-ig_1 a^{\dagger}a+\sqrt{\gamma_1}b_{1,\mathrm{in}},  \label{Eq:db1}\\
\dot{b_2}=&&-\left(\frac{\gamma_2}{2}+i\omega_2\right)b_2-i\lambda e^{-i\theta}b_1-ig_2 a^{\dagger}a+\sqrt{\gamma_2}b_{2,\mathrm{in}}, \label{Eq:db2}
\end{eqnarray}
where the total decay rate of the cavity $\kappa=\kappa_i+\kappa_e$ includes the intrinsic decay rate $\kappa_i$ and the external decay rate $\kappa_e$ because of the coupling to the microwave feedline, and $\gamma_k$ is the damping rate of the $k$th mechanical resonator. $a_{\mathrm{in}}$ and $b_{k,\mathrm{in}}$ represent the input quantum noise of the cavity and mechanical modes with zero mean values.
The steady-state solutions to Eqs. (\ref{Eq:da})-(\ref{Eq:db2}) can be obtained by setting the time derivatives to be zero, which are given by
\begin{eqnarray}
&&\alpha=\frac{\sqrt{\kappa_e}\varepsilon_L}{\frac{\kappa}{2}+i \Delta},\nonumber\\
&&\beta_1=-\frac{ig_1(\gamma_2/2+i\omega_2)+\lambda e^{i\theta}g_2}{(\gamma_1/2+i\omega_1)(\gamma_2/2+i\omega_2)+\lambda^2}|\alpha|^2,\nonumber\\
&&\beta_2=-\frac{ig_2(\gamma_1/2+i\omega_1)+\lambda e^{-i\theta}g_1}{(\gamma_1/2+i\omega_1)(\gamma_2/2+i\omega_2)+\lambda^2}|\alpha|^2,
\end{eqnarray}
where $\Delta=\Delta_c+g_1(\beta_1^*+\beta_1)+g_2(\beta_2^*+\beta_2)$ is the effective cavity-driving field detuning including the radiation pressure effects. Eqs. (\ref{Eq:da})-(\ref{Eq:db2}) can be linearized by writing each operator as the sum of its steady-state solution and a small fluctuation, i.e., $a=\alpha+\delta a$, $b_1=\beta_1+\delta b_1$, and $b_2=\beta_2+\delta b_2$. Subsequently, we have
\begin{eqnarray}
\delta\dot{a}=&&-\left(\frac{\kappa}{2}+i\Delta\right)-iG_1(\delta b_1^{\dagger}+\delta b_1)-iG_2(\delta b_2^{\dagger}+\delta b_2)\nonumber\\&&+\sqrt{\kappa_e}a_{\mathrm{in},e}+\sqrt{\kappa_i}a_{\mathrm{in},i}, \label{Eq:dda}\\
\delta\dot{b_1}=&&-\left(\frac{\gamma_1}{2}+i\omega_1\right)\delta b_1-i\lambda e^{i\theta}\delta b_2-i(G_1^*\delta a+G_1\delta a^{\dagger})\nonumber\\&&+\sqrt{\gamma_1}b_{1,\mathrm{in}}, \label{Eq:ddb1}\\
\delta\dot{b_2}=&&-\left(\frac{\gamma_2}{2}+i\omega_2\right)\delta b_2-i\lambda e^{-i\theta}\delta b_1-i(G_2^*\delta a+G_2 \delta a^{\dagger})\nonumber\\&&+\sqrt{\gamma_2}b_{2,\mathrm{in}},\label{Eq:ddb2}
\end{eqnarray}
where $G_1=g_1\alpha$ and $G_2=g_2\alpha$ are the effective optomechanical coupling strengths.

In this work, we mainly consider the cavity is driven nearly on the red sideband of the mechanical resonators, i.e., $\Delta\approx\omega_m=(\omega_1+\omega_2)/2$. Eqs. (\ref{Eq:dda})-(\ref{Eq:ddb2}) can be moved into another interaction picture by introducing the slowly moving operators with tildes, i.e., $\delta a=\delta\tilde{a}e^{-i\Delta t}$, $\delta b_1=\delta\tilde{b_1}e^{-i\omega_m t}$, and $\delta b_2=\delta\tilde{b_2}e^{-i\omega_m t}$. In the limit of $\omega_m\gg (G_{1,2},\kappa)$, the rotating wave approximation (RWA) can be invoked, and we can obtain the following equations:
\begin{eqnarray}
\dot{a}=&&-\frac{\kappa}{2} a-iG_1b_1 e^{-i\Delta_m t}-iG_2 b_2 e^{-i\Delta_m t} +\sqrt{\kappa_e}a_{\mathrm{in},e}+\sqrt{\kappa_i}a_{\mathrm{in},i},\nonumber\\\label{Eq:dta}\\
\dot{b_1}=&&-\left(\frac{\gamma_1}{2}+i\Omega\right)b_1-i\lambda e^{i\theta}b_2-iG_1^*ae^{i\Delta_m t}+\sqrt{\gamma_1}b_{1,\mathrm{in}},\label{Eq:dtb1}\\
\dot{b_2}=&&-\left(\frac{\gamma_2}{2}-i\Omega\right)b_2-i\lambda e^{-i\theta}b_1-iG_2^*ae^{i\Delta_mt}+\sqrt{\gamma_2}b_{2,\mathrm{in}},\label{Eq:dtb2}
\end{eqnarray}
where $\Delta_m=\omega_m-\Delta$, $\Omega=(\omega_1-\omega_2)/2$, and we have replaced the symbol $\delta\tilde{o}~(o=a,b_1,b_2)$ with $o$ for simplicity.
If $\Delta_m=0$, the linearized Hamiltonian of the system can be given by
\begin{eqnarray}
H_L=&&\hbar\Omega b_1^{\dagger}b_1-\hbar\Omega b_2^{\dagger}b_2+\hbar\lambda (e^{i\theta}b_1^{\dagger}b_2+e^{-i\theta}b_1 b_2^{\dagger})\nonumber\\&&+\hbar(G_1a^{\dagger}b_1+G_1^* a b_1^{\dagger})+\hbar(G_2 a^{\dagger}b_2+G_2^* ab_2^{\dagger}).\label{Eq:Hlin}
\end{eqnarray}
Without loss of generality, we assume that the optomechanical coupling strengths $G_1$ and $G_2$ are positive real numbers, and the phase $\theta$ can be viewed as a gauge phase in the loop formed by the cavity and mechanical modes (see details in Appendix \ref{app:phase}).

When the condition $\kappa\gg (G_{1,2},\gamma_{1,2})$ is satisfied, the cavity field in Eqs. (\ref{Eq:dta})-(\ref{Eq:dtb2}) can be adiabatically eliminated (see Appendix \ref{app:adiabatic}), and we obtain the effective Hamiltonian for the two mechanical resonators
\begin{equation}
H_{\mathrm{eff}}/\hbar=\left(
\begin{array}{cc}
\Omega-i\left(\frac{\gamma}{2}+\Gamma\right) & \lambda e^{i\theta}-i\Gamma\\
\lambda e^{-i\theta}-i\Gamma & -\Omega-i\left(\frac{\gamma}{2}+\Gamma\right)\\
\end{array}
\right), \label{Heff5}
\end{equation}
where we have assumed that $\Delta_m=0$, $\gamma_1=\gamma_2=\gamma$, $G_1=G_2=G$, and $\Gamma=2G^2/\kappa.$
The eigenvalues of the Hamiltonian (\ref{Heff5}) are given by
\begin{equation}
\omega_{\pm}=-i\left(\frac{\gamma}{2}+\Gamma\right)\pm\sqrt{\Omega^2+\lambda^2-\Gamma^2
-2i\Gamma\lambda\cos\theta},\label{Eq:eig}
\end{equation}
where the real parts of $\omega_{\pm}$ correspond to the resonance frequency of the mechanical eigenmodes, and the imaginary parts represent their damping rates.
If $\lambda=0$, it is easy to check that non-Hermitian Hamiltonian (\ref{Heff5}) is anti-$\mathcal{PT}$ symmetric with exceptional point at $\Gamma=\Omega$, which results from the dissipative coupling induced by the cavity. In the presence of the direct coupling between the two mechanical resonators, the eigenvalues in Eq.~(\ref{Eq:eig}) is modified and the exceptional point is shifted to $\Gamma=\sqrt{\Omega^2+\lambda^2}$ when $\theta=(2n+1)\pi/2$ with $n$ being an integer. Notable that the anti-$\mathcal{PT}$ symmetric coupling is a pure imaginary number with a fixed phase $\pi$/2, compared to the general $\mathcal{PT}$ symmetric coupling with a tunable phase $\theta$ here. Thus, the phase-match condition is defined as $\theta=(2n+1)(\pi/2)$ with $n\in Z$.

\begin{figure}
\centering
\includegraphics[width=8.6cm]{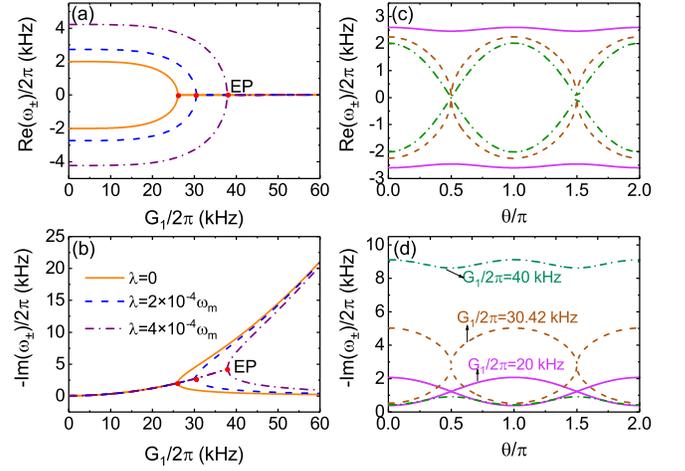}
\caption{(a) Real part Re[$\omega_{\pm}$] and (b) imaginary part -Im[$\omega_{\pm}$] of the eigenvalues as functions of the optomechanical coupling strength $G_1/2\pi$ with $\theta=\pi/2$ for different values of $\lambda$. (c) Re[$\omega_{\pm}$] and (d) -Im[$\omega_{\pm}$] versus the phase $\theta$ with $\lambda=2\times10^{-4}\omega_m$ for different values of $G_1$. The other parameters are $\omega_1/2\pi=9.285$ MHz, $\omega_2/2\pi=9.289$ MHz, $\gamma_1/2\pi=100$ Hz, $\gamma_2/2\pi=90$ Hz, $\kappa_e/2\pi=410$ kHz, $\kappa_i/2\pi=268$ kHz, $\omega_m=(\omega_1+\omega_2)/2$, $\Delta=\omega_m$, and we keep $G_1=G_2$ throughout this work.}
\label{Fig2}
\end{figure}
To demonstrate how the gauge field and phase will affect the energy-level evaluation in this optomechanical system, we choose the experimentally realizable parameters from a recent work~\cite{KorppiPRA}: $\omega_1/2\pi=9.285$ MHz, $\omega_2/2\pi=9.289$ MHz, $\gamma_1/2\pi=100$ Hz, $\gamma_2/2\pi=90$ Hz, $\kappa_e/2\pi=410$ kHz, and $\kappa_i/2\pi=268$ kHz. In Fig.~\ref{Fig2}, we plot the real and imaginary parts of the eigenvalues as functions of the (a) optomechanical coupling strength $G_1/2\pi$ and (b) phase $\theta$. If the two mechanical modes are not coupled directly ($\lambda=0$), the solid lines in Figs. \ref{Fig2}(a)-\ref{Fig2}(b) show that the system can exhibit the anti-$\mathcal{PT}$ symmetry by modulating the coupling strength $G_1$. At lower value of $G_1$, the splitting between the two real parts of the eigenvalues are approximately equal to the resonance frequency difference between the two mechanical modes, and the imaginary parts (damping rates) are nearly the same. With increasing the optomechanical coupling strength $G_1$, the two mechanical modes are coupled stronger via their common interaction with the cavity field, however, the splitting between the two real parts Re$(\omega_{\pm})$ gets smaller, i.e., energy-level attracted together. More specially, there exists the EPs where both the real and imaginary parts of the eigenvalues coalesce. At higher value of $G_1$, the eigenvalues becomes purely imaginary and the system works in the unbroken anti-$\mathcal{PT}$-symmetric regime. Such counterintuitive energy-attraction attributed to anti-$\mathcal{PT}$ symmetry broken differs to normal mode splitting observed with $\mathcal{PT}$ symmetry broken, when increasing mode couplings. As also shown in Fig.~\ref{Fig2}(a), if the two mechanical modes are coupled directly, the eigenvalues will be modified. When the coupling strength $\lambda$ between the two mechanical modes is increased from 0 to $4\times10^{-4}\omega_m$ with fixed phase $\theta=\pi/2$, the splitting between the two real parts Re$(\omega_{\pm})$ becomes larger and the EP (red dot) is shifted to a higher value of $G_1$.

The phase dependence of the eigenvalues is calculated and plotted in Figs.~\ref{Fig2}(c)-\ref{Fig2}(d). The splitting between the two real (imaginary) parts of the eigenvalues reaches the maximum at $\theta=n\pi$ and the minimum at $\theta=(2n+1)\pi/2(n\in Z)$, i.e., phase-math is satisfied. By increasing the optomechanical coupling strength $G_1/2\pi$ from 20 kHz to 40 kHz, the splitting between the two real parts gets smaller but the splitting between the two imaginary parts becomes larger. Especially when $G_1/2\pi=30.42$ kHz, both the real and imaginary parts coalesce at the phase-match points e.g., $n=(0,1)$ corresponding to $\theta=(0.5,1.5)\pi$. The EP in Figs.~\ref{Fig2}(c)-\ref{Fig2}(d) with $\theta=0.5\pi$ corresponds to the EP with $\lambda=2\times10^{-4}\omega_m$ in Figs. \ref{Fig2}(a)-\ref{Fig2}(b). We emphasize that the counterintuitive energy-level-attraction originates from the destructive interference between the $\mathcal{PT}$- and anti-$\mathcal{PT}$-symmetric coupling paths from the gauge field. Furthermore, the exceptional points periodically appear when phase-match is satisfied.

\section{Observations of energy attraction through transmission and output spectra} \label{sec:transmission}
Based on the optomechanical interactions, cavity field supports a versatile platform to observe the mechanical energy-level evolutions. We now study the transmission and output spectrum of the cavity in this section. We define the vectors $\mu=(a,b_1,b_2)^{\mathrm{T}}$ for the system operators, $\mu_{\mathrm{in}}=(a_{\mathrm{in},e},a_{\mathrm{in},i},b_{1,\mathrm{in}},b_{2,\mathrm{in}})^{\mathrm{T}}$ for the input operators, then Eqs. (\ref{Eq:dta})-(\ref{Eq:dtb2}) can be written in the matrix form
\begin{equation}
\dot{\mu}=-M\mu+L\mu_{\mathrm{in}}, \label{Eq:mu}
\end{equation}
where the coefficient matrix
\begin{equation}
M=
\left(
\begin{array}{ccc}
\kappa/2 & iG_1 & iG_2\\
iG_1^* & \gamma_1/2+i\Omega & i\lambda e^{i\theta}\\
iG_2^* & i\lambda e^{-i\theta} & \gamma_2/2-i\Omega\\
\end{array}
\right), \label{Eq:M}
\end{equation}
and
\begin{equation}
L=
\left(
\begin{array}{cccc}
\sqrt{\kappa_e} & \sqrt{\kappa_i} & 0 & 0\\
0 & 0 & \sqrt{\gamma_1}& 0\\
0 & 0 & 0 & \sqrt{\gamma_2}\\
\end{array}
\right). \label{Eq:L}
\end{equation}

Introducing the Fourier transform
\begin{eqnarray}
&&o(\omega)=\int_{-\infty}^{\infty} o(t)e^{i\omega t}dt,\nonumber\\
&&o^{\dagger}(\omega)=\int_{-\infty}^{\infty}o^{\dagger}(t) e^{i\omega t}dt,
\end{eqnarray}
the solution to Eq. (\ref{Eq:mu}) is then given by
\begin{equation}
\mu(\omega)=(M-i\omega)^{-1}L\mu_\mathrm{in}(\omega).
\end{equation}
According to the input-output relation $\mu_\mathrm{out}(\omega)=\mu_\mathrm{in}(\omega)-L^\mathrm{T}\mu(\omega)$~\cite{ClerkRMP} with $\mu_{\mathrm{out}}(\omega)=(a_{\mathrm{out},e}, a_{\mathrm{out},i}, b_{1,\mathrm{out}},b_{2,\mathrm{out}})^\mathrm{T}$ being the vector for the output operators, we can obtain $\mu_\mathrm{out}(\omega)=S(\omega)\mu_{\mathrm{in}}(\omega)$ with the transmission matrix
\begin{equation}
S(\omega)=I_{4\times4}-L^{\mathrm{T}}(M-i\omega)^{-1}L.
\end{equation}
It is easy to derive that
\begin{eqnarray}
&&S_{11}(\omega)=1-\kappa_e(\Gamma_1\Gamma_2+\lambda^2)/d(\omega),\\
&&S_{12}(\omega)=-\sqrt{\kappa_e \kappa_i}(\Gamma_1\Gamma_2+\lambda^2)/d(\omega),\\
&&S_{13}(\omega)=\sqrt{\kappa_e\gamma_1}(iG_1\Gamma_2+G_2\lambda e^{-i\theta})/d(\omega),\\
&&S_{14}(\omega)=\sqrt{\kappa_e\gamma_2}(iG_2\Gamma_1+G_1\lambda e^{i\theta})/d(\omega),
\end{eqnarray}
where $d(\omega)=\Gamma_c\Gamma_1\Gamma_2+\Gamma_1 G_2^2+\Gamma_2 G_1^2+\Gamma_c\lambda^2-2i\lambda G_1G_2\cos\theta$ with $\Gamma_c=\frac{\kappa}{2}-i\omega$, $\Gamma_1=\frac{\gamma_1}{2}+i(\Omega-\omega)$, and $\Gamma_2=\frac{\gamma_2}{2}-i(\Omega+\omega)$.
The input quantum noises of the cavity and mechanical modes satisfy the correlation function
\begin{eqnarray}
&&\langle a_{\mathrm{in}}(\omega)a_\mathrm{in}^{\dagger}(\omega')\rangle=2\pi\delta(\omega+\omega'),\\
&&\langle b_{k,\mathrm{in}}(\omega)b_{k,\mathrm{in}}^{\dagger}(\omega')\rangle=2\pi(n_k+1)\delta(\omega+\omega'),\\
&&\langle b_{k,\mathrm{in}}^{\dagger}(\omega)b_{k,\mathrm{in}}(\omega')\rangle=2\pi n_k\delta(\omega+\omega'),
\end{eqnarray}
where $n_k=1/[\exp(\hbar\omega_k/k_\mathrm{B}T_e)-1]$ corresponds to the thermal phonon occupation of the mechanical mode at the environment temperature $T_e$.
Therefore, the output spectrum of the cavity is given by
\begin{eqnarray}
S_{\mathrm{out}}=&&\frac{1}{2}\int\frac{d\omega'}{2\pi}\langle a_{\mathrm{out},e}(\omega)a_{\mathrm{out},e}^{\dagger}(\omega')
+a_{\mathrm{out},e}^{\dagger}(\omega')a_{\mathrm{out},e}(\omega)\rangle\nonumber\\
=&&\frac{1}{2}|S_{11}|^2+\frac{1}{2}|S_{12}|^2+(n_1+\frac{1}{2})|S_{13}(\omega)|^2\nonumber\\&&+(n_2+\frac{1}{2})|S_{14}(\omega)|^2,
\end{eqnarray}

\begin{figure}
\centering
\includegraphics[width=8.6cm]{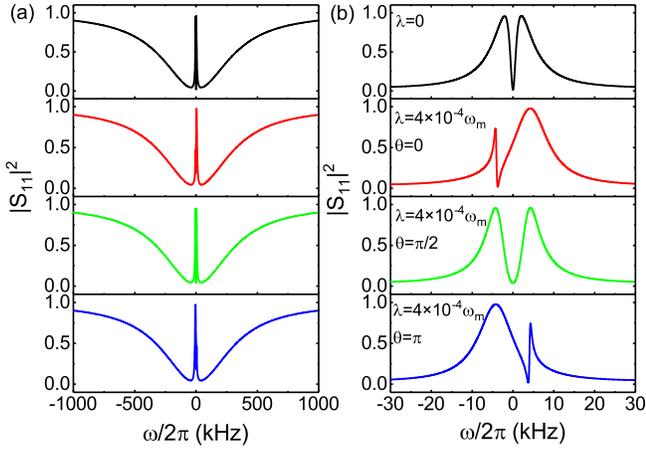}
\caption{Transmission probability $|S_{11}|^2$ as a function of the probe detuning $\omega/2\pi$ for different values of the phase $\theta$. Panel (b) is the detail of panel (a) in the vicinity of $\omega=0$. The other parameters are the same as those in Fig. \ref{Fig2} except $G_1/2\pi=30$ kHz.}
\label{Fig3}
\end{figure}
In Fig. \ref{Fig3}, the transmission probability $|S_{11}|^2$ is plotted versus the probe detuning $\omega/2\pi$ when the phase $\theta$ varies. Figure \ref{Fig3}(a) shows the transmission probability $|S_{11}|^2$ in a wide range of frequency, and the details of the peaks around $\omega=0$ can be clearly seen from Fig. \ref{Fig3}(b). If only considering the anti-$\mathcal{PT}$-symmetric coupling between the two mechanical modes (i.e., $\lambda=0$), two symmetric transparency peaks locate at $\omega=\pm\Omega$, which can be referred to as double optomechanically induced transparency and explained by the interference effect~\cite{JiangCPB,KorppiPRA}. When the two mechanical modes are coupled directly, the system forms a closed interaction loop and the phase effect becomes important. If $\lambda=4\times10^{-4}\omega_m$ and $\theta=\pi/2$, the two transparency windows are still symmetric, but the splitting between the two peaks is broadened. The position of the peaks are approximately determined by $\omega=\pm\sqrt{\Omega^2+\lambda^2}$ at low value of $G_1$, as shown in Fig. \ref{Fig2}(a). However, if the phase $\theta$ is tuned to be $0$ or $\pi$, the transmission spectrum exhibits an asymmetric Fano line shape with a narrow dip and a broad transparency window induced by the interference effect. Under the phononic gauge field, it is notable that symmetric transparency only exist in the case of phase-match. As an orthogonal perspective, the position of the transmission dip can be controlled by the phase $\theta$, but the analytical expression is too cumbersome to be given here.

\begin{figure}
\centering
\includegraphics[width=8.6cm]{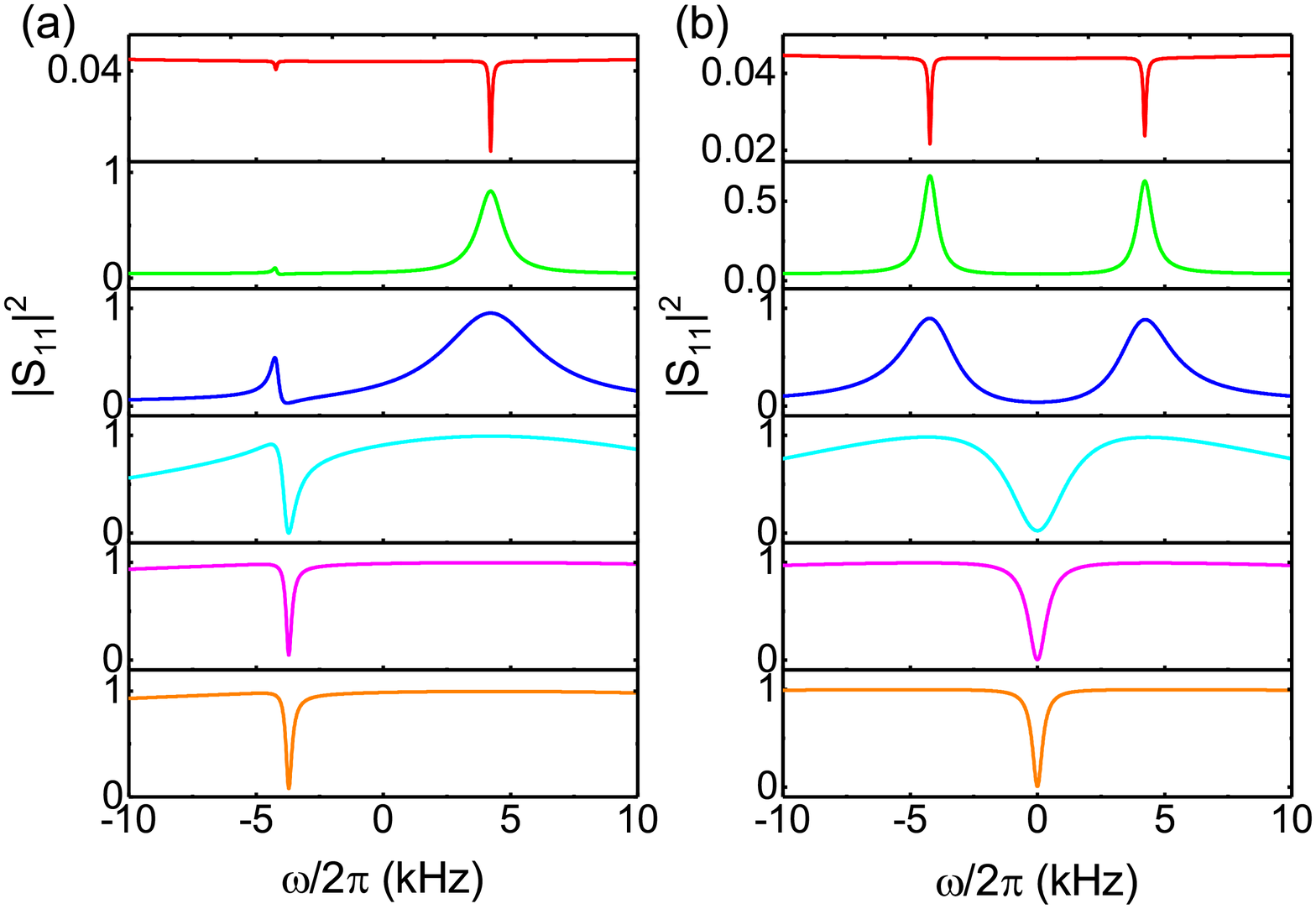}
\caption{Transmission probability $|S_{11}|^2$ as a function of the probe detuning $\omega/2\pi$ with (a) $\theta=0$ and (b) $\theta=\pi/2$ for $G_1/2\pi=(1,10,20,50,90,130)$ kHz from top to bottom.  The other parameters are the same as those in Fig. \ref{Fig2} except $\lambda=4\times10^{-4}\omega_m$.}
\label{Fig4}
\end{figure}

When the power of the driving field is increased, the effective optomechanical coupling strength will be enhanced. Figure~\ref{Fig4} plots the transmission probability $|S_{11}|^2$ as functions the probe detuning $\omega/2\pi$ for a series of the value $G_1$ with (a) $\theta=0$ and (b) $\theta=\pi/2$. At $G_1/2\pi=1$ kHz, the anti-$\mathcal{PT}$-symmetric coupling between the two mechanical modes mediated by the radiation pressure is weak, and the transmission spectrum exhibits two narrow dips around the cavity center due to the anti-Stokes scattering process. When the optomechanical coupling strength is enhanced to be $G_1/2\pi=10$ kHz, the two dips turn into two peaks, which results from the destructive interference between the anti-Stokes field and the probe field. Note that the two peaks are asymmetric at $\theta=0$ but the two peaks are symmetric at $\theta=\pi/2$ (i.e., phase-matched with $n=0$). With further increasing the coupling strength $G_1$, the linewidth of the peaks are broadened. Under phase-match, e.g., $\theta=\pi/2$, the linewidth is given by the effective mechanical damping rate $\gamma_{k,\mathrm{eff}}=\gamma_{k}+\gamma_{k,\mathrm{opt}}$ with $\gamma_{k,\mathrm{opt}}=4G_k^2/\kappa$. When the condition $\gamma_{k,\mathrm{eff}}>|\omega_1-\omega_2|$ is satisfied~\cite{KorppiPRA} around the EP, the individual mechanical modes have large spectral overlap and the two transmission peaks gradually merge into a single dip in the cavity center. The linewidth of the transmission dip becomes smaller with the increase of the coupling strength $G_1$, which can also seen from the lower branch of the curve above the EP in Fig. \ref{Fig2}(b). At $\theta=0$, the transmission dip is evolved from the asymmetric Fano line shape with zero transmission probability on the left of Fig.~\ref{Fig4}(a). Figure~\ref{Fig4} clearly shows that mechanical energy attraction happens when the phase-match condition is satisfied.

\begin{figure}
\centering
\includegraphics[width=8.6cm]{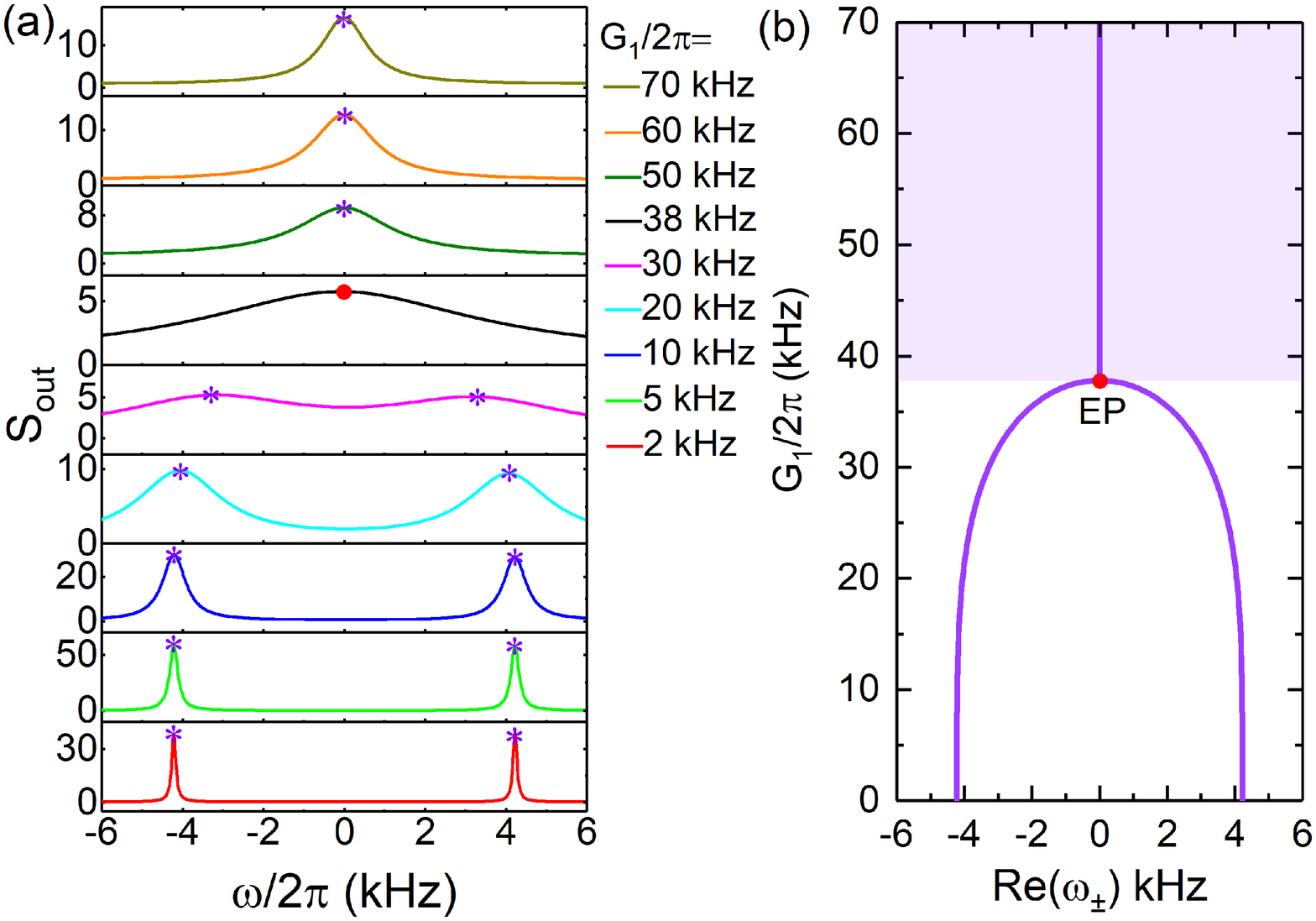}
\caption{(a) Cavity output spectrum versus the probe detuning $\omega/2\pi$ for varying the optomechanical coupling strength from $G_1/2\pi=2$ kHz (bottom curve) to $G_1/2\pi=70$ kHz (top curve). (b) The real part Re[$\omega_{\pm}$] of the eigenvalue versus the coupling strength $G_1/2\pi$. The other parameters are the same as those in Fig. \ref{Fig4} except $n_1=n_2=100$ and $\theta=\pi/2$. Red dots mark the EPs under the phase-match.}
\label{Fig5}
\end{figure}
Furthermore, the exceptional point in this optomechanical system can be more intuitively demonstrated by measuring the cavity output spectrum, as shown in Fig. \ref{Fig5}. At low value of the optomechanical coupling strength $G_1$, two symmetric peaks can be observed in the output spectrum, which corresponds to the two mechanical eigenmodes. When the strength $G_1/2\pi$ is increased from 2 kHz to 30 kHz, the splitting between the two peaks becomes smaller but the linewidth gets larger. At $G_1/2\pi=38$ kHz, i.e., the exceptional point, the two peaks merge into a single peak in the cavity center, which can be viewed as the level attraction between the two mechanical eigenmodes due to enhanced optomechanical coupling. The position of the single peak remains the same but the linewidth decreases with further increasing the coupling strength $G_1$. In particular, the envelope of the peak values in Fig. \ref{Fig5}(a) forms the curve for the real part Re$(\omega_{\pm})$ of the eigenvalues shown in Fig.~\ref{Fig5}(b).

\section{Heating-resistant ground-state cooling}\label{sec:cooling}
Taking the dissipations into consideration, the evolution for the density matrix of the optomechanical system is governed by the quantum master equation
\begin{eqnarray}
\dot{\rho}=&&-\frac{i}{\hbar}[H_L,\rho]+\kappa\mathcal{D}[a]\rho+\gamma_1(n_1+1)\mathcal{D}[b_1]\rho
+\gamma_1 n_1\mathcal{D}[b_1^{\dagger}]\rho\nonumber\\&&+\gamma_2(n_2+1)\mathcal{D}[b_2]\rho
+\gamma_2 n_2\mathcal{D}[b_2^{\dagger}]\rho,
\end{eqnarray}
where $\mathcal{D}[o]\rho=o\rho o^{\dagger}-\frac{1}{2}o^{\dagger}o\rho-\frac{1}{2}\rho o^{\dagger}o$ is the standard Lindblad superoperator for the dissipations of the cavity and mechanical modes, and the Hamiltonian $H_L$ is given by Eq. (\ref{Eq:Hlin}). According to $\langle\dot{o}\rangle=Tr\{o\dot{\rho}\}$, we can obtain the time evolution of the second-order moments, $\langle a^{\dagger}a\rangle$, $\bar{n}_1=\langle b_1^{\dagger}b_1\rangle$, $\bar{n}_2=\langle b_2^{\dagger}b_2\rangle$, $\langle a^{\dagger} b_1\rangle$, $\langle a^{\dagger} b_2\rangle$, and $\langle b_1^{\dagger} b_2\rangle$. The differential equations are given by
\begin{widetext}
\begin{eqnarray}
&&\frac{d}{dt}\langle a^{\dagger}a\rangle=-\kappa\langle a^{\dagger}a\rangle-i\left(G_1\langle a^{\dagger}b_1\rangle+G_2\langle a^{\dagger}b_2\rangle-G_1\langle a^{\dagger}b_1\rangle^*-G_2\langle a^{\dagger}b_2\rangle^*\right),\label{Eq:m1}\\
&&\frac{d}{dt}\langle b_1^{\dagger}b_1\rangle=-\gamma_1\langle b_1^{\dagger}b_1\rangle+i\left(G_1\langle a^{\dagger} b_1\rangle-G_1\langle a^{\dagger} b_1\rangle^*-\lambda e^{i\theta}\langle b_1^{\dagger}b_2\rangle+\lambda e^{-i\theta}\langle b_1^\dagger b_2\rangle^*\right)+\gamma_1 n_1,\\
&&\frac{d}{dt}\langle b_2^{\dagger}b_2\rangle=-\gamma_2\langle b_2^{\dagger} b_2\rangle+i\left(G_2\langle a^{\dagger} b_2\rangle-G_2\langle a^{\dagger} b_2\rangle^*+\lambda e^{i\theta}\langle b_1^{\dagger}b_2\rangle-\lambda e^{-i\theta}\langle b_1^{\dagger}b_2\rangle^*\right)+\gamma_2 n_2,\\
&&\frac{d}{dt}\langle a^{\dagger} b_1\rangle=\left(-\frac{\kappa+\gamma_1}{2}-i\Omega\right)\langle a^{\dagger} b_1\rangle+i\left(G_1\langle b_1^{\dagger}b_1\rangle+G_2\langle b_1^{\dagger}b_2\rangle^*-G_1\langle a^{\dagger}a\rangle-\lambda e^{i\theta}\langle a^{\dagger}b_2\rangle\right),\\
&&\frac{d}{dt}\langle a^{\dagger} b_2\rangle=\left(-\frac{\kappa+\gamma_2}{2}+i\Omega\right)\langle a^{\dagger} b_2\rangle+i\left(G_1\langle b_1^{\dagger}b_2\rangle+G_2\langle b_2^{\dagger}b_2\rangle-G_2\langle a^{\dagger}a\rangle-\lambda e^{-i\theta}\langle a^{\dagger}b_1\rangle\right),\\
&&\frac{d}{dt}\langle b_1^{\dagger}b_2\rangle=\left(-\frac{\gamma_1+\gamma_2}{2}+2i\Omega\right)\langle b_1^{\dagger}b_2\rangle+i\left(G_1\langle a^{\dagger}b_2\rangle-G_2\langle a^{\dagger} b_1\rangle^*-\lambda e^{-i\theta}\langle b_1^{\dagger}b_1\rangle+\lambda e^{-i\theta}\langle b_2^{\dagger}b_2\rangle\right).\label{Eq:m6}
\end{eqnarray}
\end{widetext}
In the steady state, all the derivatives in Eqs. (\ref{Eq:m1})-(\ref{Eq:m6}) equal to zero. Under the condition of $\kappa>G\gg\{\lambda,\Omega,\gamma_{1,2}\}$ with $G_1=G_2=G$ and $\gamma_m=(\gamma_1+\gamma_2)/2$, the final average phonon numbers can be obtained as
\begin{eqnarray}
&&\bar{n}_1^f\approx\frac{2(\Gamma^2-\lambda^2)\gamma_1n_1+2(\Gamma\mp\lambda)^2\gamma_2n_2}
{[\gamma_m^2+4(\lambda^2+\Omega^2+\Gamma\gamma_m)](2\Gamma+\gamma_m)}+\frac{\gamma_1n_1}{2\Gamma+\gamma_m},\label{Eq:n1f}\\
&&\bar{n}_2^f\approx\frac{2(\Gamma\pm\lambda)^2\gamma_1n_1+2(\Gamma^2-\lambda^2)\gamma_2n_2}
{[\gamma_m^2+4(\lambda^2+\Omega^2+\Gamma\gamma_m)](2\Gamma+\gamma_m)}+\frac{\gamma_2n_2}{2\Gamma+\gamma_m},\label{Eq:n2f}
\end{eqnarray}
where the upper (lower) sign in ``$\mp$'' and ``$\pm$'' corresponds to $\theta=\pi/2(3\pi/2)$.

\begin{figure}
\centering
\includegraphics[width=8.6cm]{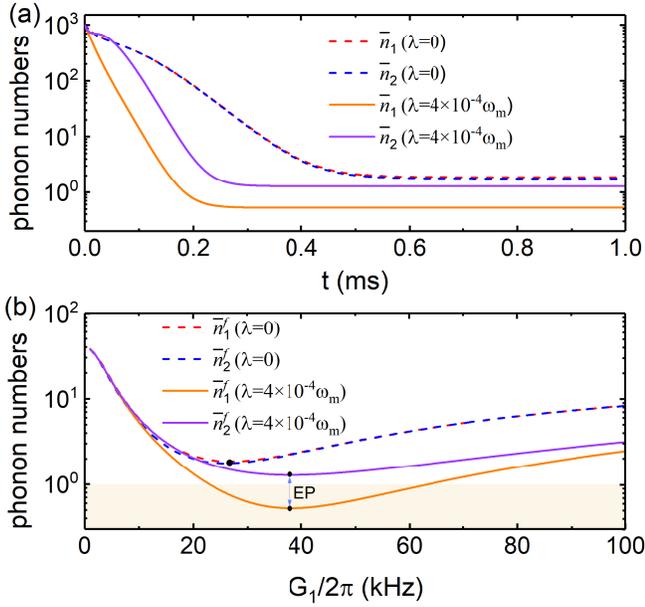}
\caption{(a) The time evolution of the average phonon numbers $\bar{n}_1$ and $\bar{n}_2$ when the mechanical-mechanical coupling is turned off ($\lambda=0, G_1/2\pi=26$ kHz, dashed lines) and on ($\lambda=4\times 10^{-4}\omega_m, \theta=\pi/2, G_1/2\pi=38$ kHz, solid lines). (b) The final average phonon numbers $\bar{n}_1^f$ and $\bar{n}_2^f$ as a function of the optomechanical coupling strength $G_1/2\pi$ when $\lambda=0$ (dashed lines) and $\lambda=4\times 10^{-4}\omega_m, \theta=\pi/2$ (solid lines). The black-solid dots located at the minimum phonon number mark the EPs. The other parameters are the same as those in Fig. \ref{Fig4} except $n_1=n_2=40$.}
\label{Fig6}
\end{figure}
We study the cooling of the mechanical resonators by numerically solving Eqs.~(\ref{Eq:m1})-(\ref{Eq:m6}). Figure~\ref{Fig6}(a) plots the time evolution of the average phonon numbers without and with the $\mathcal{PT}$-symmetric direct coupling between the mechanical resonators. Without $\mathcal{PT}$-symmetric couplings, i.e., $\lambda=0$, the final steady-state phonon occupations are almost the same for these two mechanical modes. For example, steady-state phonon numbers are $\bar{n}_1\approx1.9$ and $\bar{n}_2\approx1.8$ with $G_1/2\pi=26$ kHz. Both these two mechanical modes are not arriving at the ground-state if the mechanical resonators just have optomechanical mediated anti-$\mathcal{PT}$-symmetric couplings. The final (steady state) average phonon numbers $\bar{n}_1^f$ and $\bar{n}_2^f$ versus the optomechanical coupling strength $G_1/2\pi$ are plotted in Fig.~\ref{Fig6}(b). Further increasing the optomechanical coupling $G_1$, the phonon numbers start to monotonically increase instead of further cooling as expected. Recall the EPs discussions [e.g., EPs in Fig.~\ref{Fig2}(a)], the system undergoes a transition into the symmetry-broken phase where the dark modes formed by these mechanical modes decouple from the cavity mode and prevent extracting energy from the dark modes through the cooling channel of the cavity mode.

We now discuss how will the phononic gauge field affect mechanical cooling. As shown in Fig~\ref{Fig6}, the two mechanical resonators can be further cooled by turning on the $\mathcal{PT}$-symmetric direct coupling, i.e., constructing a phononic gauge. It is shown that $\bar{n}_1\approx0.5$ and $\bar{n}_2\approx1.3$ in the steady state at $\lambda=4\times10^{-4}\omega_m$, $\theta=\pi/2$, and $G_1/2\pi=38$ kHz. We emphasize that with such phononic gauge field and under the phase-match (e.g., $n$=1, $\theta=\pi/2$): (i) both these two mechanical resonators hold smaller steady-state phonon numbers; (ii) with the increase of the value $G_1$, the final average phonon numbers decrease monotonically and reach the minimum at the EP; (iii) the final steady-state phonon occupations are quite different and the ground-state cooling for mechanical 1 is realized. Further increasing $G_1$ until crossing the EPs, the final average phonon numbers starts to increase again. This unexpected sabotage for the cooling also attributes to the mechanical dark mode formed in the symmetry-broken regime. It is notable that the EPs shifted to the right when considering the direct mechanical coupling, reducing the final steady-state phonon number and then the ground-state cooling is achieved.




\begin{figure}
\centering
\includegraphics[width=8.6cm]{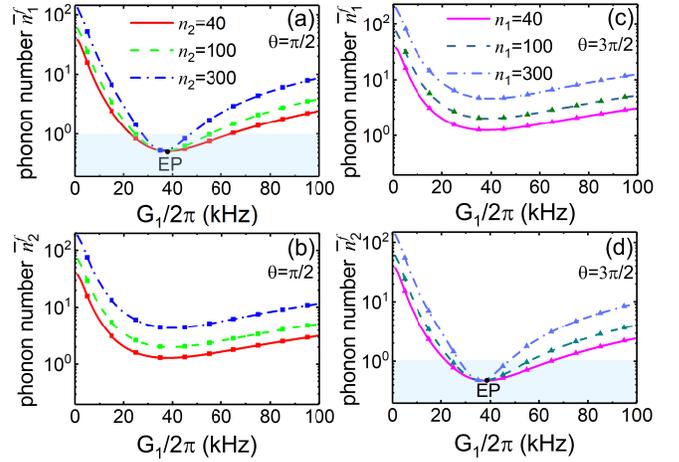}
\caption{The final average phonon numbers $\bar{n}_1^f$ and $\bar{n}_2^f$ versus the optomechanical coupling strength $G_1/2\pi$ for different thermal phonon occupations with (a)-(b) $\theta=\pi/2$ and (c)-(d) $\theta=3\pi/2$. The curves and the symbols correspond to the numerical results based on Eqs. (\ref{Eq:m1})-(\ref{Eq:m6}) and the analytical results of Eqs. (\ref{Eq:n1f})-(\ref{Eq:n2f}), respectively. The other parameters are the same as those in Fig. \ref{Fig4} except $n_1=40$ in (a)-(b) and $n_2=40$ in (c)-(d).}
\label{Fig7}
\end{figure}

Finally, we discuss how will the gauge phase affect the thermal phonon transport and the mechanical cooling performance. Figure~\ref{Fig7}(a) and~\ref{Fig7}(b) plots the final average phonon numbers $\bar{n}_1^f$ and $\bar{n}_2^f$ as a function of the coupling strength $G_1/2\pi$ with $\theta=\pi/2$ (i.e., $n=0$) and $n_1=40$ for different values of $n_2$. It is shown that both the final average phonon numbers $\bar{n}_1^f$ and $\bar{n}_2^f$ increase with the thermal phonon occupation $n_2$. However, the minimum value of $\bar{n}_1^f$ keeps constant at the exceptional point when the thermal phonon occupation $n_2$ increases, which demonstrates the robustness of the cooling limit in the first mechanical resonator against the thermal noise of the second mechanical resonator at the EP. It can also be seen from Eq.~(\ref{Eq:n1f}) that $\bar{n}_1^f$ is independent on the $n_2$ when $\theta=\pi/2$ and $\Gamma=\lambda$, which is consistent with the numerical results. If the phase is tuned to be $\theta=3\pi/2$ (i.e., $n=1$), Fig.~\ref{Fig7}(d) shows that the second mechanical resonator can be cooled to the ground state around the EP, which is also robust against the thermal noise of the first mechanical resonator. This phenomenon is closely related with the nonreciprocal phonon transfer which becomes ideally unidirectional at the EPs.~Recalling Eqs.~(\ref{Eq:n1f}) and~(\ref{Eq:n2f}), we reveal that phonon including its thermal noise transports from mechanical 2 (1) to 1 (2) is blocked when $n$ is even (odd) number. Our work revealed that the thermal energy transfer can be controlled by tuning the gauge phase, which supports a promising route to overpass the notorious heating limitations. Note that for the resolved sideband regime ($\kappa<\omega_{1,2}$) under consideration, Stokes scattering due to the finite cavity linewidth limits the final average phonon number to $\bar{n}_{ba}=(\kappa/4\omega_m)^2\approx3\times10^{-4}$~\cite{Wilson-Rae,Marquardt}, which has been neglected here.


\section{Conclusion}\label{sec:conclusion}
In summary, we have investigated the energy-level evolution and the cooling of the mechanical resonators under a phase tunable phononic gauge field. By adiabatically eliminating the cavity mode, we revealed that the effective coupling between two mechanical modes can be purely imaginary where the anti-$\mathcal{PT}$-symmetry and phase-match are correspondingly defined. The transmission spectrum then exhibits the asymmetric Fano line shape or double optomechanically induced transparency by modulating the gauge phase. Subsequently, the counterintuitive energy-level attraction accompanied by periodical EPs are observed under the phase-match condition. Besides from the transmission spectrum, we proposed that such energy-level-attraction and the caused EPs can be more intuitively observed in the cavity output power spectrum where the mechanical eigenvalues correspond to the peaks.

The gauge field and phase also greatly affects the phonon transport. Especially for mechanical cooling, the average phonon occupation number becomes minimum at these EPs and the mechanical resonator can arrives at the ground state. Moreover, destructive quantum interference happens within the gauge field and then the phonon transport becomes nonreciprocal and even ideally unidirectional at the EPs under phase match. The thermal blockade direction is switchable and controlled by the gauge phase. Finally, we proposed a heating-resistant ground-state cooling based on the nonreciprocal phonon transport. Towards the quantum regime of macroscopic mechanical resonators, most optomechanical systems are ultimately limited by their intrinsic cavity or mechanical heating originated from the material defects, photothermal conversion, or phase noise. The proposed heating-resistant ground-state cooling happens at the EPs, which is closely related to the topological Riemann-sheet. Our work may motivate more explorations towards overpassing the notorious heating limitations, e.g., through the topological protection by encircling the EPs~\cite{Zhong2018,Doppler2016}.

\section*{ACKNOWLEDGMENTS}
C.~J. was supported by the Natural Science Foundation of China (NSFC) under Grant No. 11874170 and the Postdoctoral Science Foundation of China under Grant No. 2017M620593. Y.L.L acknowledges the financial support of the Natural Science Foundation of China (NSFC) under Grant No. 12004044.

\appendix
\section{GAUGE PHASE}\label{app:phase}
In Sec. \ref{sec:system}, we assume that the optomechanical coupling strengths $G_1$ and $G_2$ are positive real numbers and $\theta$ represents a gauge phase. Here, we explain this by redefining the operators $b_1$ and $b_2$. In general, the coupling strengths $G_1=g_1\alpha=|G_1|e^{i\theta_1}$ and $G_2=g_2\alpha=|G_2|e^{i\theta_2}$ with $\theta_1=\theta_2$. We refine the operators as
\begin{eqnarray}
b_1\rightarrow b_1 e^{-i\theta_1},b_2\rightarrow b_2 e^{-i\theta_2},
\end{eqnarray}
then the linearized Hamiltonian (\ref{Eq:Hlin}) becomes
\begin{eqnarray}
H_{L}=&&\hbar\Omega b_1^\dagger b_1-\hbar\Omega b_2^{\dagger}b_2\nonumber\\&&+\hbar\lambda\left[e^{i(\theta+\theta_1-\theta_2)}b_1^\dagger b_2+e^{-i(\theta+\theta_1-\theta_2)}b_1 b_2^{\dagger}\right]\nonumber\\&&+\hbar(|G_1|a^\dagger b_1+|G_1|a b_1^\dagger)+\hbar(|G_2|a^\dagger b_2+a b_2^\dagger).
\end{eqnarray}
Therefore, by redefining the operators, the phase $\theta=\theta+\theta_1-\theta_2$ can be treated as a gauge phase in the loop formed by the modes $a,b_1,b_2$, and the coupling strengths $G_{1,2}\rightarrow|G_{1,2}|$, becoming positive real numbers.

\section{ADIABATIC ELIMINATION}\label{app:adiabatic}
In order to obtain the effective Hamiltonian about the mechanical resonators, we neglect the noise terms for simplicity.
According to Eq. (\ref{Eq:dta}), we can obtain the formal solution of $a$ as
\begin{eqnarray}
a(t)=&&-iG_1\int_0^t dt' b_1(t')e^{-i\Delta_m t'}e^{-\frac{\kappa}{2}(t-t')}\nonumber\\&&-iG_2\int_0^t dt' b_2(t')e^{-i\Delta_m t'}e^{-\frac{\kappa}{2}(t-t')}. \label{Eq:Sa1}
\end{eqnarray}
If the decay rate of the cavity is large enough and satisfies $\kappa\gg\gamma_1,\gamma_2$, then the changes of mode $b_1$ and mode $b_2$ are small within the rage of integration of the cavity mode $a$. Therefore, we can set $b_1(t')\approx b_1(t)$, $b_2(t')\approx b_2(t)$, and then take them out of the integral in Eq. (\ref{Eq:Sa1}) to obtain
\begin{eqnarray}
a(t)=&&-iG_1 b_1(t)\int_0^t dt'e^{-i\Delta_m t'}e^{-\frac{\kappa}{2}(t-t')}\nonumber\\&&-iG_2 b_2(t)\int_0^t dt' e^{-i\Delta_m t'}e^{-\frac{\kappa}{2}(t-t')}\nonumber\\
=&&-iG_1b_1(t)\frac{e^{-i\Delta_m t}}{\frac{\kappa}{2}-i\Delta_m}-iG_2 b_2(t)\frac{e^{-i\Delta_m t}}{\frac{\kappa}{2}-i\Delta_m}. \label{Eq:Sa2}
\end{eqnarray}
Substituting Eq. (\ref{Eq:Sa2}) into Eqs. (\ref{Eq:dtb1})-(\ref{Eq:dtb2}), we can adiabatically eliminate the cavity mode $a$ to obtain the following equations of motion for the mechanical mode $b_1$ and $b_2$:
\begin{eqnarray}
\dot{b_1}=&&-\left(\frac{\gamma_1}{2}+i\Omega+\frac{|G_1|^2}{\frac{\kappa}{2}-i\Delta_m}\right)b_1
-\left(\frac{G_1^*G_2}{\frac{\kappa}{2}-i\Delta_m}+i\lambda e^{i\theta}\right)b_2,\label{Eq:b1e}\nonumber\\
\end{eqnarray}
\begin{eqnarray}
\dot{b_2}=&&-\left(\frac{\gamma_2}{2}-i\Omega+\frac{|G_2|^2}{\frac{\kappa}{2}-i\Delta_m}\right)b_2
-\left(\frac{G_1 G_2^*}{\frac{\kappa}{2}-i\Delta_m}+i\lambda e^{-i\theta}\right)b_1,\label{Eq:b2e}\nonumber\\
\end{eqnarray}
Eqs. (\ref{Eq:b1e})-(\ref{Eq:b2e}) can be written in the matrix form as
\begin{eqnarray}
&&i\frac{d}{dt}\left(
\begin{array}{cc}b_1\\b_2\\
\end{array}\right)\nonumber\\&&=
\left(
\begin{array}{cc}
\Omega-i\left(\frac{\gamma}{2}+\frac{|G_1|^2}{\kappa/2-i\Delta_m}\right) & \lambda e^{i\theta}-i\frac{G_1^*G_2}{\kappa/2-i\Delta_m}\\
\lambda e^{-i\theta}-i\frac{G_1 G_2^*}{\kappa/2-i\Delta_m} & -\Omega-i\left(\frac{\gamma_2}{2}+\frac{|G_2|^2}{\kappa/2-i\Delta_m}\right)\\
\end{array}
\right)\left(
\begin{array}{cc}b_1\\b_2\\
\end{array}\right). \nonumber\\\label{Heff1}
\end{eqnarray}
Therefore, the effective Hamiltonian for the two mechanical resonators can be given by
\begin{equation}
H_{\mathrm{eff}}/\hbar=\left(
\begin{array}{cc}
\Omega-i\left(\frac{\gamma_1}{2}+\frac{|G_1|^2}{\kappa/2-i\Delta_m}\right) & \lambda e^{i\theta}-i\frac{G_1^*G_2}{\kappa/2-i\Delta_m}\\
\lambda e^{-i\theta}-i\frac{G_1 G_2^*}{\kappa/2-i\Delta_m} & -\Omega-i\left(\frac{\gamma_2}{2}+\frac{|G_2|^2}{\kappa/2-i\Delta_m}\right)\\
\end{array}
\right) \label{Heff2}
\end{equation}
If the damping rates of the two mechanical resonators are the same $(\gamma_1=\gamma_2=\gamma)$, the effective optomechanical coupling strengths are the same and real $(G_1=G_2=G)$, and the cavity is driven close to the red sideband of the mechanical resonator $(\Delta_m= 0)$, the effective Hamiltonian Eq. (\ref{Heff2}) is reduced to
\begin{equation}
H_{\mathrm{eff}}/\hbar=\left(
\begin{array}{cc}
\Omega-i\left(\frac{\gamma}{2}+\Gamma\right) & \lambda e^{i\theta}-i\Gamma\\
\lambda e^{-i\theta}-i\Gamma & -\Omega-i\left(\frac{\gamma}{2}+\Gamma\right)\\
\end{array}
\right), \label{Heff3}
\end{equation}
where $\Gamma=2G^2/\kappa.$
Equivalently, the Hamiltonian Eq. (\ref{Heff3}) can be written as
\begin{eqnarray}
H_{\mathrm{eff}}/\hbar=&&\left[\Omega-i\left(\frac{\gamma}{2}+\Gamma\right)\right]b_1^{\dagger}b_1-\left[
\Omega+i\left(\frac{\gamma}{2}+\Gamma\right)\right]b_2^{\dagger}b_2\nonumber\\&&+\left(\lambda e^{-i\theta}-i\Gamma\right)b_1^{\dagger}b_2+\left(\lambda e^{i\theta}-i\Gamma\right)b_1b_2^{\dagger}.
\end{eqnarray}
We note that the coupling between the two mechanical resonators can be classified into two categories. The term $H_{I1}=\lambda e^{-i\theta}b_1^{\dagger}b_2+\lambda e^{i\theta}b_1b_2^{\dagger}$ represents the $\mathcal{PT}$-symmetric coupling since $(\mathcal{PT})H_{I1}(\mathcal{PT})^{-1}=H_{I1}$ under the parity $\mathcal{P}$ (i.e., $b_1\leftrightarrow b_2$) and time-reversal $\mathcal{T}$ (i.e., $i\leftrightarrow -i$) operations. The term $H_{I2}=-i\Gamma b_1^{\dagger}b_2-i\Gamma b_1 b_2^{\dagger}$ corresponds to the anti-$\mathcal{PT}$-symmetric coupling since $(\mathcal{PT})H_{I2}(\mathcal{PT})^{-1}=-H_{I2}$ under the parity $\mathcal{P}$ and time-reversal $\mathcal{T}$ operations.

\end{document}